\documentclass{PoS}

\def\lsi{\raise0.3ex\hbox{$<$\kern-0.75em\raise-1.1ex\hbox{$\sim$}}}
\def\gsi{\raise0.3ex\hbox{$>$\kern-0.75em\raise-1.1ex\hbox{$\sim$}}}
\def\backder{\raise1.4ex\hbox{$\leftarrow$\kern-0.75em\raise-1.4ex\hbox{$\partial$}}}
\def\forder{\raise1.4ex\hbox{$\rightarrow$\kern-0.8em\raise-1.4ex\hbox{$\partial$}}}

\newcommand{\gsim}{\mathop{\gsi}}
\newcommand{\C}{{\kern+.25em\sf{C}\kern-.45em\sf{{\small{I}}} \kern+.45em\kern-.25em}}
\newcommand{\R}{{\kern+.25em\sf{R}\kern-.78em\sf{I} \kern+.78em\kern-.25em}}

\newcommand{\beq}{\begin{equation}}
\newcommand{\eeq}{\end{equation}}

\newcommand{\vep}{\varepsilon}

\title{Schwinger model simulations with dynamical overlap fermions}

\ShortTitle{Schwinger model with dynamical overlap fermions}

\author{\speaker{Wolfgang Bietenholz}%
\\
NIC/DESY Zeuthen, Platanenallee 6 \\
D-15738 Zeuthen, Germany \\
        E-mail: \email{bietenho@ifh.de}}
\author{Stanislav Shcheredin\\
Fakult\"{a}t f\"{u}r Physik, Univerisit\"{a}t Bielefeld \\
D-33615 Bielefeld, Germany\\
E-mail: \email{shchered@physik.hu-berlin.de}}
\author{Jan Volkholz\\
Institut f\"{u}r Physik, Humboldt-Universit\"{a}t zu Berlin \\
Newtonstr. 15, D-12489 Berlin, Germany\\
E-mail: \email{volkholz@physik.hu-berlin.de}}

\abstract{We present simulation results for the 2-flavour 
Schwinger model with dynamical overlap fermions. In particular
we apply the overlap hypercube operator at seven
light fermion masses. In each case we collect sizable statistics
in the topological sectors 0 and 1. Since the chiral condensate 
$\Sigma$ vanishes in the chiral limit, we observe
densities for the microscopic Dirac spectrum, which have
not been addressed yet by Random Matrix Theory (RMT).
Nevertheless, by confronting the averages of the lowest eigenvalues
in different topological sectors with chiral RMT in unitary ensemble 
we obtain --- for the very light fermion masses --- values
for $\Sigma$ that follow closely the analytical predictions 
in the continuum.}

\FullConference{The XXV International Symposium on Lattice Field
Theory\\
                 July 30 - August 4 2007\\
                 Regensburg, Germany}

\begin{document}

\section{The Schwinger model}

The Schwinger model \cite{schwingmod}
describes Dirac fermions in $d=2$,
interacting through a $U(1)$ gauge field. 
In the Euclidean plane the Lagrangian reads
\beq
{\cal L} (\bar \psi , \psi , A_{\mu}) =
\bar \psi (x) \Big[ \gamma_{\mu} ( i \partial_{\mu} + g A_{\mu} ) + m \Big]
\psi (x) + \frac{1}{2} F_{\mu \nu}(x) F_{\mu \nu}(x) \ .
\eeq
This is a popular toy model for QCD --- 
for instance it is endowed with confinement.
As a qualitative difference,
however, there is no spontaneous chiral symmetry breaking.
For $N_{f}$ degenerate flavours of mass $m$
the chiral condensate is given by\footnote{For $N_{f}=1$
the non-vanishing value $\Sigma (0) = (e^{\gamma}/ 2 \pi^{3/2}) \, g
\simeq 0.16 \, g$ originates from an axial anomaly and 
therefore from explicit chiral symmetry breaking 
(hence there is no contradiction to the Mermin-Wagner theorem).}
\beq
\Sigma (m) \equiv - \langle \bar \psi \, \psi \rangle \propto 
\Big( \frac{m^{N_{f}-1}}{\beta} \Big)^{1/(N_{f}+1)} \qquad
( \beta = 1 /g^{2}) \ .
\eeq
Here we consider $N_{f}=2$. In this case, there are analytical evaluations
(using low energy assumptions) for the above proportionality constant in the 
case of light fermions ($m \ll 1 / \sqrt{\beta}$),
\beq  \label{Sigmam}
\Sigma (m) = \left\{ \begin{array}{lcr}
0.372 \ ( m / \beta ) ^{1/3} && \hspace*{1cm} \cite{HHI} \\
0.388 \ ( m / \beta ) ^{1/3} && \hspace*{1cm} \cite{Smilga} 
\end{array} \right.
\eeq

\section{Lattice formulation}

We investigate the lattice formulation with compact link variables
$U_{\mu ,x}\in U(1)$, and with the plaquette gauge action. 
For the fermions we employ an overlap hypercube fermion 
(overlap-HF) Dirac operator of the form (in lattice units)
\beq
D_{\rm ovHF}(m) = \Big( 1 - \frac{m}{2} \Big) \, D_{\rm ovHF}^{(0)} 
+ m \ , 
\quad D_{\rm ovHF}^{(0)} = 1 + ( D_{\rm HF} -1) / 
\sqrt{( D_{\rm HF}^{\dagger} -1)( D_{\rm HF} -1)} \ .
\eeq
$D_{\rm ovHF}^{(0)}$ obeys the (simplest) Ginsparg-Wilson relation.
%$\{ D_{\rm ovHF}^{(0)} , \gamma_{5} \} = D_{\rm ovHF}^{(0)} \gamma_{5} 
%D_{\rm ovHF}^{(0)}
Unlike the standard overlap operator with a
Wilson kernel \cite{Neu}, we insert the truncated perfect
hypercube fermion operator $D_{\rm HF}$ \cite{WBEPJC}.
It involves couplings to nearest neighbour sites, and over 
plaquette diagonals (in the latter case gauging
averages over the shortest lattice paths). 
By construction this kernel is approximately chiral already,
and the overlap formula amounts to a correction that renders
chirality exact.\footnote{To be precise, we use the 
chirally-optimised hypercube fermion (CO-HF) of Ref.\ \cite{WBIH}. 
This is optimal for our algorithm to be described in Section 3.}

The overlap-HF has been applied in quenched QCD \cite{qQCD}, and
the HF was also used dynamically in finite temperature QCD
\cite{StaniEdwin}.
In the 2-flavour Schwinger model $D_{\rm ovHF}$ has been first
simulated with quenched re-weighted configurations \cite{WBIH,Quatsch1}.
Compared to the standard overlap operator there is
some computational overhead in the kernel, but $D_{\rm ovHF}$ 
has the following virtues \cite{WBIH}:
\vspace*{-1mm}
\begin{itemize}
\item Faster convergence in the polynomial evaluation of $D_{\rm ovHF}$.
Moreover the limitation to the use of low polynomials also 
improves the numerical stability.
\vspace*{-1mm}
\item Higher degree of locality and approximate rotation symmetry.
\vspace*{-1mm}
\item Improved scaling behaviour.
\vspace*{-1mm}
\end{itemize}
All these virtues are based on the similarity
of the kernel to the overlap operator \cite{WBEPJC},
\begin{equation}  \label{approxHF}
D_{\rm ovHF} \approx D_{\rm HF} \ .
\end{equation}

\section{The simulation}

%\vspace*{-1mm}

Here we report on HMC simulations, which  are also facilitated
by the property (\ref{approxHF}).
Our algorithmic concept follows the simplified HMC force 
for improved staggered fermions of the HF-type \cite{Dilg}.
The fermionic force of the standard HMC algorithm
\beq
\bar \psi Q_{\rm ovHF}^{-1} \Big[ Q_{\rm ovHF}^{-1} 
\frac{\partial Q_{\rm ovHF}}{\partial A_{x,\mu}} +
\frac{\partial Q_{\rm ovHF}}{\partial A_{x,\mu}} Q_{\rm ovHF}^{-1} 
\Big] Q_{\rm ovHF}^{-1} \psi \ ,
\eeq
with the Hermitian operator $Q_{\rm ovHF} = \gamma_{5} D_{\rm ovHF}$ ,
is simplified to
\beq  \label{simple}
\bar \psi Q_{{\rm ovHF},\vep }^{-1} \Big[ 
Q_{{\rm ovHF},\vep}^{-1} 
\frac{\partial Q_{\rm HF}}{\partial A_{x,\mu}} +
\frac{\partial Q_{\rm HF}}{\partial A_{x,\mu}} 
Q_{{\rm ovHF},\vep}^{-1}
\Big] Q_{{\rm ovHF},\vep }^{-1} \psi \ .
\eeq
$Q_{{\rm ovHF},\vep }$ approximates $Q_{\rm ovHF}$ to a 
moderate (absolute) precision of $\vep = 10^{-5} \, $. 
This approximation is useful and cheap thanks to 
relation (\ref{approxHF}) (which does not apply for the
standard overlap operator).
The Metropolis accept/reject step uses $Q_{\rm ovHF}$
to machine precision ($10^{-16}$), which renders the algorithm exact.
Our first experience at $\beta =5$ on 
a $16 \times 16$ lattice, with trajectory
length $\tau = 1/8 = 20 \cdot \Delta \tau$, was reported in
Ref.\ \cite{Lat06}. Applying the Sexton-Weingarten integration 
scheme \cite{SexWein}, we have meanwhile a compelling confirmation
of acceptance rates in the range $0.3 \dots 0.5$ for
the masses $m = 0.01 \dots 0.24$. %see Figure \ref{accfig}.
Acceptance rates for the special case
$Q_{{\rm ovHF},\vep} \equiv Q_{{\rm HF}}$ were also given
in Ref.\ \cite{Quatsch2}. 
Our results show a remarkable stability in $m$ down to
very light fermions. This holds for the total computing
effort as well; note that the magnitude of the leading non-zero Dirac 
eigenvalue stabilises due to the finite size.
%However, for the extremest masses used, $m =0.01$ und $0.24$,
%we occasionally had to divide the trajectory length by 2
%due to technical problems (lack of convergence).
In Ref.\ \cite{Lat06} we demonstrated that reversibility 
holds to a good precision. 
%\begin{figure}
%\begin{center}
%\includegraphics[angle=0,width=.5\linewidth]{acceptance_lat06.eps}
%\end{center}
%\caption{The acceptance rate as a function of the fermion mass.
%{\tt [The errors should shrink, $m=0.01$ should be added and hopefully
%also a larger volume, e.g. $20 \times 20$ or $24 \times 24$.]} }
%\label{accfig}
%\end{figure}
The degree of locality is stable in $m$ and strongly improved,
even compared to the free standard overlap fermion, where the
couplings decay as $\exp (-r)$ ($r$ being the taxi driver distance 
between source and sink). For the free overlap-HF this 
decay is accelerated to $\exp (- 1.5 r)$. At $\beta =5$ it slows down
only slightly to $\exp (- 1.45 r)$, with hardly any dependence
on the masses that we investigated \cite{Lat06}.
%\ref{locfig}.
%\begin{figure}
%\begin{center}
%\includegraphics[angle=270,width=.5\linewidth]{overlaploc2d.eps}
%\hspace*{-7mm}
%\includegraphics[angle=270,width=.5\linewidth]{loc2d.eps}
%\end{center}
%\caption{The degree of locality for free overlap fermions
%(on the left) and interaction overlap-HFs at $\beta=5$ (on the right).
%$f(r)$ is the maximal impact of a unit source over a taxi driver
%distance $r$ \cite{HJL}.}
%\label{locfig}
%\end{figure}

%\vspace*{-1mm}

\section{Results}

%\vspace*{-1mm}

In view of the $\epsilon$-regime,
we simulated at 7 fermion masses and collected data
in the sectors with topological charge $\nu =0$ and $|\nu | = 1$
(index of $D_{\rm ovHF}$).
%For the measurements we considered configurations separated
%by a total trajectory length $1/2$. 
The corresponding statistics and the mean values of the leading 
non-zero eigenvalue $\lambda_{1}$ of the Dirac operator 
--- \ stereographically projected
onto a line --- \ are given in Table \ref{tabstat}. 
%The errors were evaluated by the jack-knife method.

\begin{table}
\centering
\begin{tabular}{|c||r|c||r|c||c||c|}
\hline
$m$ & $\nu =0$ & & $|\nu | =1$ & &
           total      & topological \\
 & & $\langle \lambda_{1, \, \nu =0} \rangle$ & & 
$\langle \lambda_{1, \, |\nu | =1} \rangle$ & statistics & transitions \\
\hline
\hline
0.01 &  2079  & 0.1328(7) &  584 & 0.1735(10) & 2663 &  3 \\  
\hline
0.03 &  1131  & 0.1311(18) &  563 & 0.1737(24) & 1668 &  2 \\
\hline
0.06 &   752  & 0.1254(24) &  711 & 0.1728(20) & 1398 &  5 \\ 
\hline
0.09 &   957  & 0.1157(22) & 546 & 0.1713(24) & 1504 &  7 \\
\hline
0.12 &   699  & 0.1082(28) & 532 & 0.1664(26) & 1505 &  8 \\
\hline
0.18 &   830  & 0.1076(28) & 609 & 0.1660(24) & 1493 & 13 \\
\hline
0.24 &   639  & 0.1096(28) & 1030 & 0.1642(18) & 1757 & 17 \\
\hline
\end{tabular}
\caption{Our statistics for seven fermion masses in the
sectors with topological charge $\nu =0$ and $|\nu | = 1$.
The (stereographically projected) leading non-zero eigenvalue 
$\lambda_{1}$ of the Dirac operator is measured separately in 
each sector.}
\label{tabstat}
\end{table}

Chiral RMT has been worked out for the case of a non-vanishing
chiral condensate $\Sigma$ in the chiral limit. 
This yields predictions for the low lying Dirac eigenvalues \cite{WGW}
in the $\epsilon$-regime, which apply well in QCD \cite{QCDq,epsQCD}.
We show in Figure \ref{cumdens} the measured cumulative densities 
in our case, in the topologically trivial sector. This is compared
to the RMT predictions for the parameters we are using
(we refer to the unitary ensemble; the corresponding formulae are 
summarised in the second work quoted in Ref.\ \cite{epsQCD}).
They converge in terms of the dimensionless
rescaled eigenvalue $\zeta_{1} = \lambda_{1} \Sigma V$
for very light or very heavy masses, where
the latter limit corresponds to the quenched case.
In the chiral limit this is obviously inapplicable in our situation. 
The plot in Figure \ref{cumdens} on the right also shows that the
shape of the density for $\zeta_{1}$ at finite mass
does not match the RMT predictions.
Instead we see a stabilisation in the eigenvalue $\lambda_{1}$ 
itself (in a fixed volume $V$). 

In this setting a total density $\rho (\lambda) \propto \lambda^{1/3}$
is consistent with eq.\ ({\ref{Sigmam}) \cite{DHNS}, and we can approximately
confirm this behaviour, see Figure \ref{totdens}. The exponent is not
singled out very precisely, but the essential observation is
the absence of a Banks-Casher type plateau in the total
eigenvalue density near 0.

\begin{figure}
\begin{center}
\includegraphics[angle=270,width=.5\linewidth]{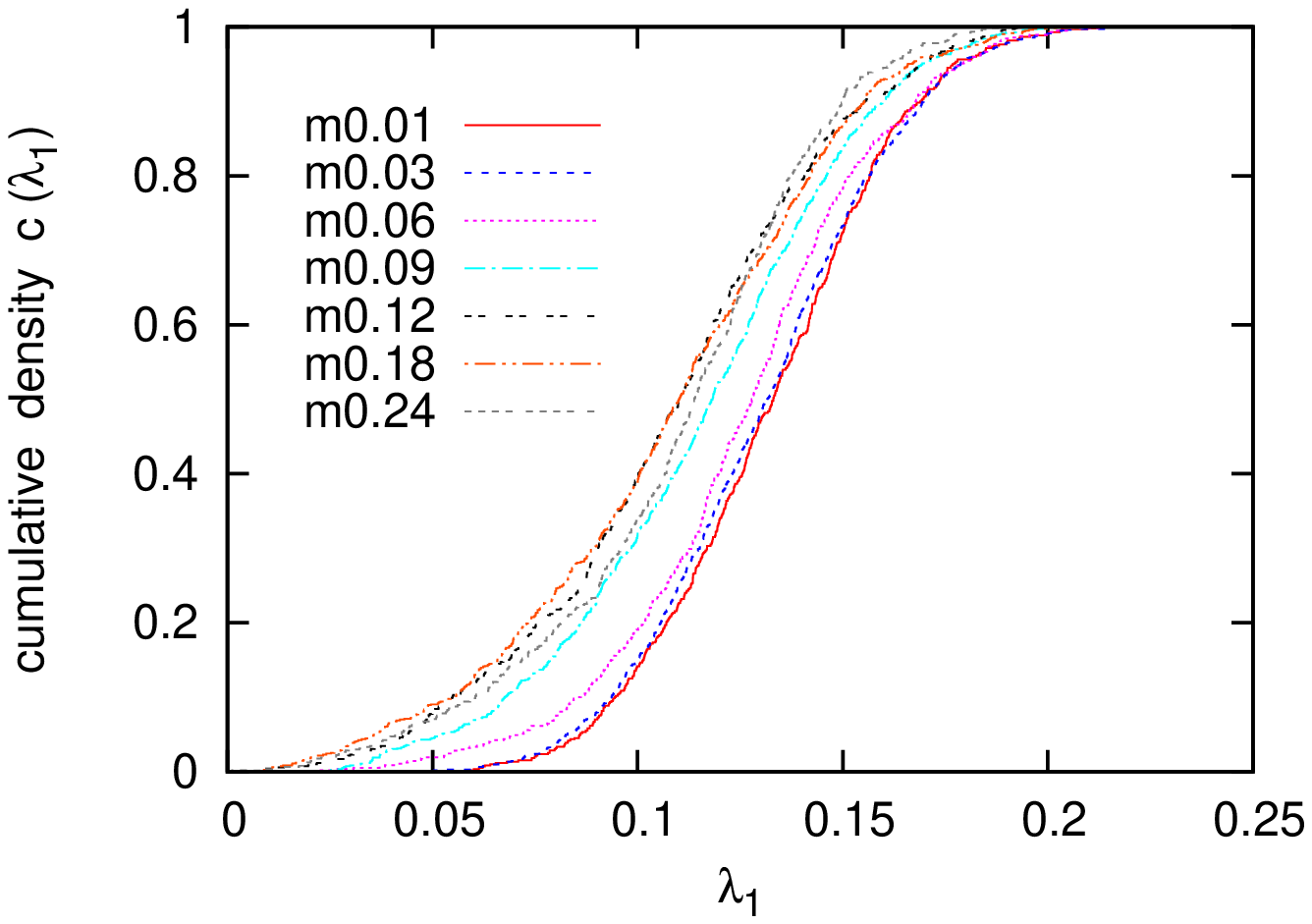}
\hspace*{-2mm}
\includegraphics[angle=270,width=.5\linewidth]{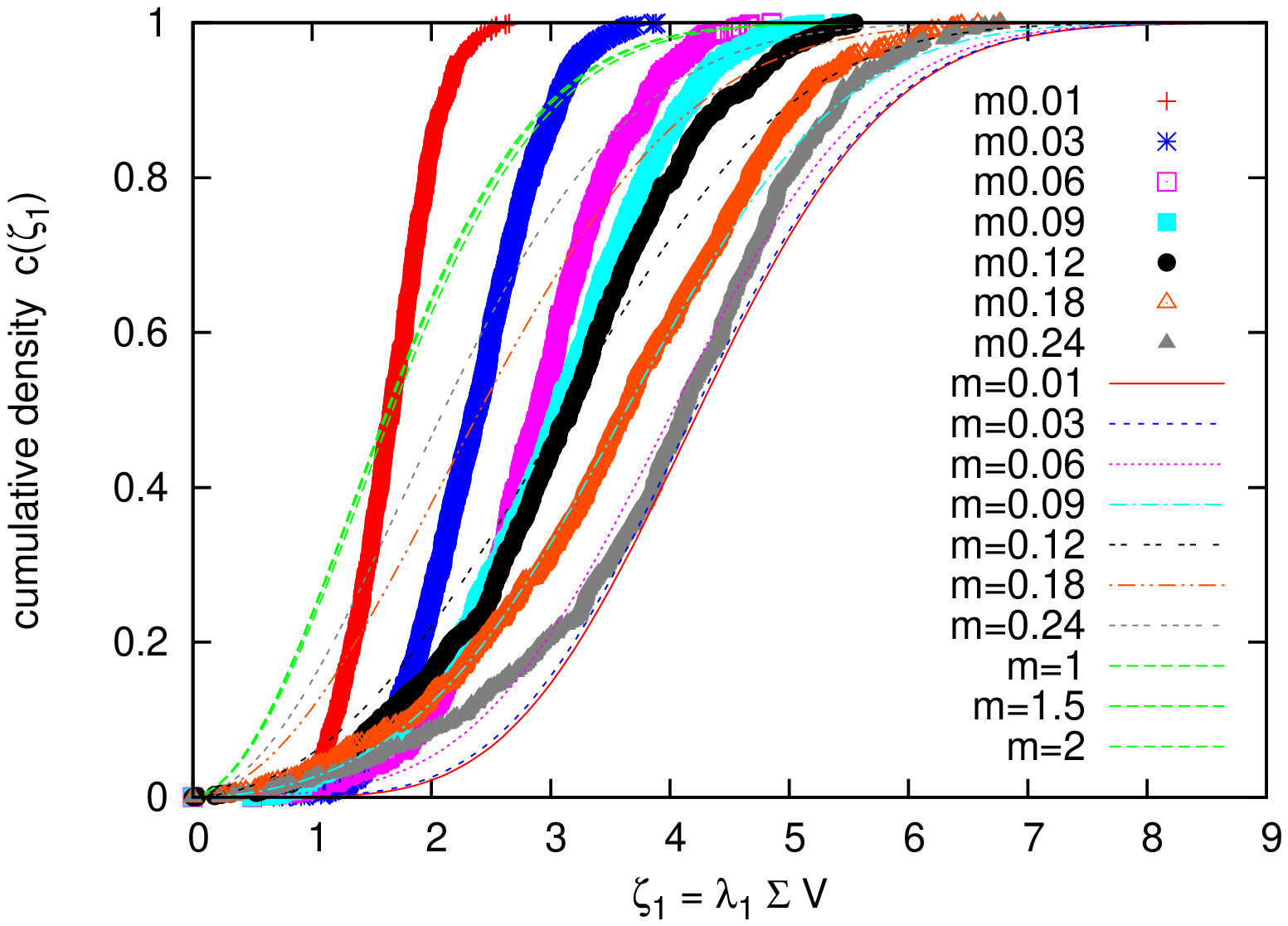}
\end{center}
\caption{On the left: the cumulative densities of the lowest Dirac 
eigenvalues for different fermion
masses. On the right: the chiral RMT prediction with the corresponding
parameters, which differs from the data
as expected. The RMT curves converge for
$\zeta_{1} \equiv \lambda_{1} \Sigma V$ in the limits $\mu \equiv
m \Sigma V \to 0$ and $\mu \to \infty \, $,
whereas in the measured data the density of $\lambda_{1}$ stabilises
in the chiral limit.}
\label{cumdens}
\end{figure}

\begin{figure}
\begin{center}
\includegraphics[angle=270,width=.49\linewidth]{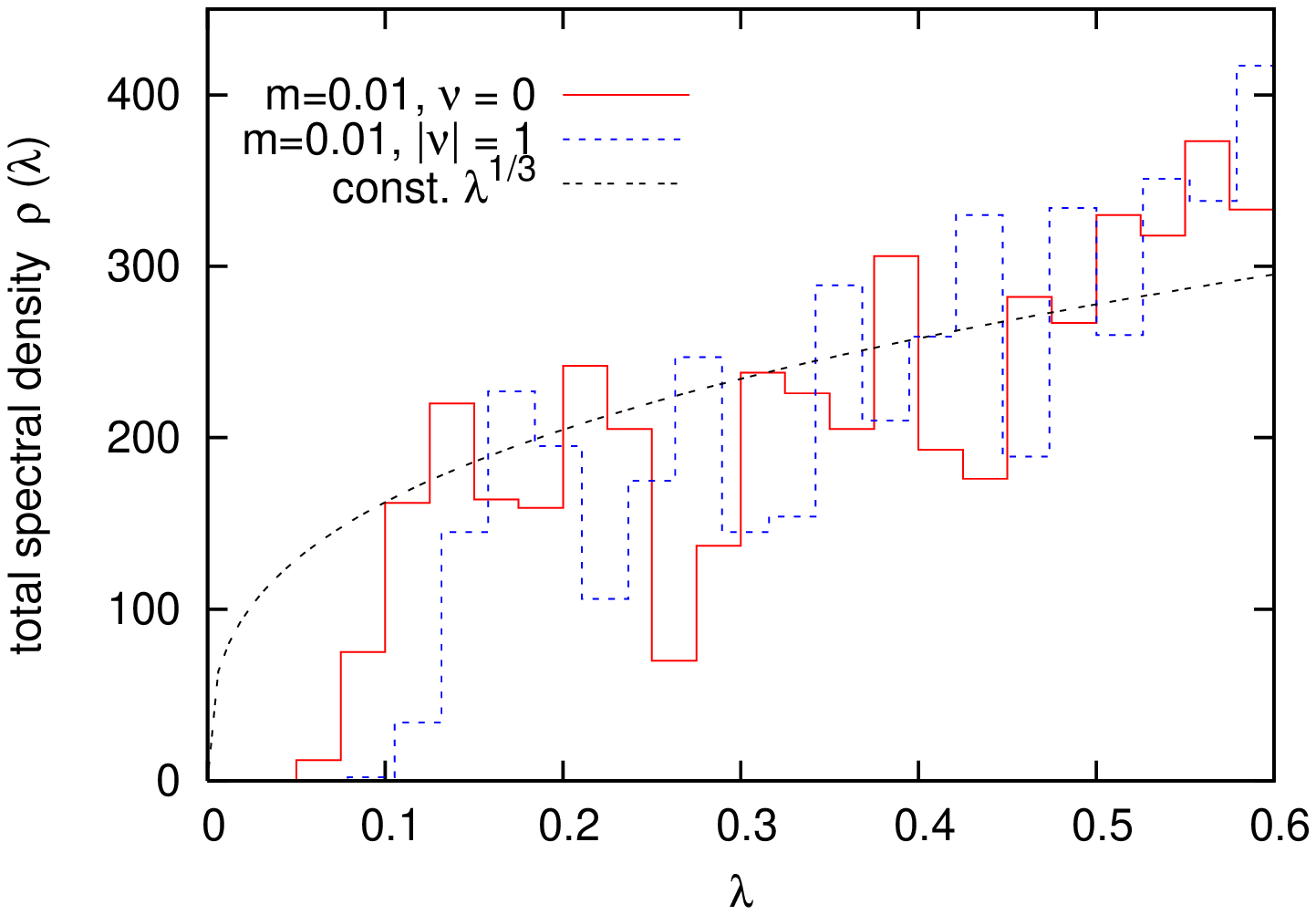}
\hspace*{-2mm}
\includegraphics[angle=270,width=.49\linewidth]{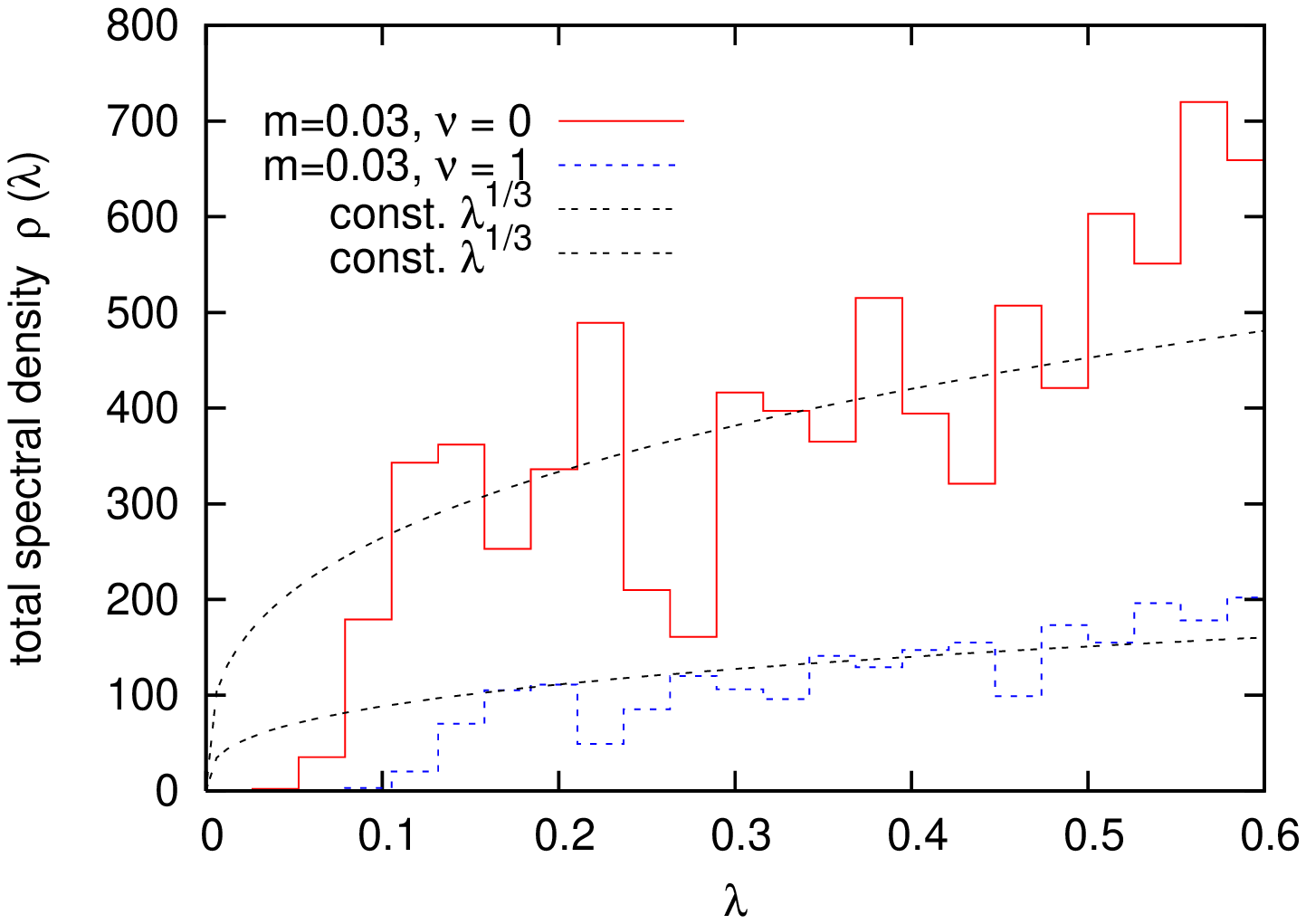}
%\hspace*{-2mm}
%\includegraphics[angle=270,width=.32\linewidth]{specdensm0.18.eps}
\end{center}
\caption{Histograms for the total eigenvalue density $\rho (\lambda)$
near zero for our two lightest
fermion masses. The data are consistent with the expected behaviour
$\rho (\lambda) \propto \lambda^{1/3}$, and we see a wiggle structure
in addition.}
\label{totdens}
\end{figure}

Nevertheless, we observed an amazing connection to chiral RMT
with respect to the {\em ratio} of $\langle \lambda_{1} \rangle$
%the mean values of the leading Dirac eigenvalues 
in the sectors with topological charge
$|\nu | =0 $ and $1$. 
%We now demonstrate and make use of this property. 
We illustrated in Ref.\ \cite{Lat06}
the chiral condensate $\Sigma$ as a function of
this ratio at various masses, according to
chiral Random  Matrix Theory \cite{WGW}.
%First we shown in Figure \ref{sigmalam} the RMT
%predictions for this eigenvalue ratio in the topological 
%sectors $0$ and $1$, cf.\ Table \ref{tabstat}.
%\begin{figure}
%\begin{center}
%\includegraphics[angle=270,width=.5\linewidth]{Sigma_lambda3.eps}
%\end{center}
%\caption{The chiral condensate $\Sigma$ as a function of
%the ratio of the leading non-zero Dirac eigenvalues
%in the topological sectors $0$ and $1$, according to 
%chiral Random  Matrix Theory \cite{WGW}.}
%\label{sigmalam}
%\end{figure}
The combination of this RMT relation with 
$\Sigma (m)$ in eq.\ (\ref{Sigmam}) enables us to eliminate the
chiral condensate and to arrive at a prediction for the ratio
$$
\frac{\langle \lambda_{1, \, | \nu | = 1} \rangle}
{\langle \lambda_{1, \, \nu = 0} \rangle} (m) \ ,
$$ 
which can be directly confronted with the
numerical data in Table \ref{tabstat}, see Figure \ref{EVratvsm}.

The simulation results reveal a significant
dynamical effect. For masses $m \gsim 0.15$ we
take a step towards the $p$-regime behaviour
(insensitivity to $\nu$) and the condition
$m \ll \beta^{-1/2}$ is not on solid grounds anymore. 
But for $m \leq 0.12$ the data match the predictions remarkably 
well (at very light masses the measured ratio tends to be just slightly
above the prediction), although the latter combines apparently
incompatible ingredients
from chiral RMT the $\epsilon$-regime and from infinite volume.
This result can be compared to a study using quenched re-weighted
configurations with the standard overlap operator \cite{DuHo}: that
study obtained $\Sigma (m \to 0) \to 0$ and a behaviour consistent
with $\Sigma \propto m^{1/3}$ at large masses, but the proportionality
constant was not reproduced and the proportionality could not be observed
at small masses. In our case, Figure \ref {EVratvsm} is sensitive to both,
the exponent and the proportionality constant in eq. (\ref{Sigmam}),
and both are confirmed well.

\begin{figure}
\vspace*{-3mm}
\begin{center}
\includegraphics[angle=270,width=.7\linewidth]{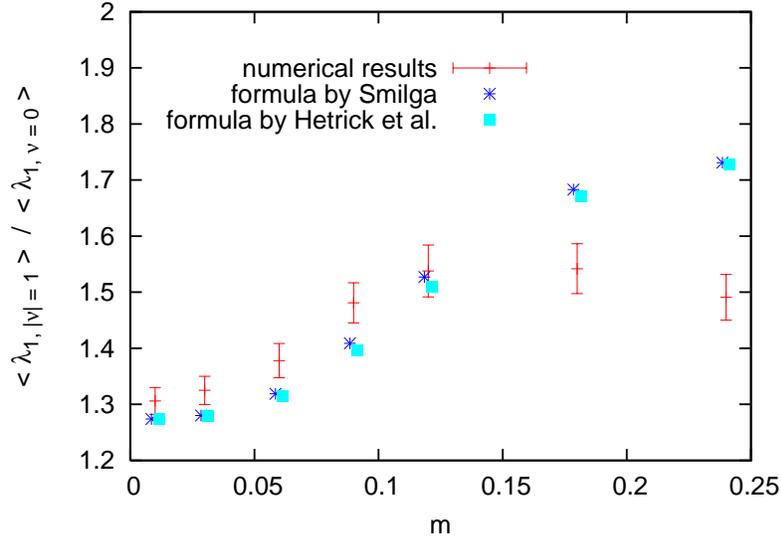}
\caption{The results for the eigenvalue ratio of the leading
non-zero Dirac eigenvalues in the topological sectors 0 and 1,
$\langle \lambda_{1, \, | \nu | = 1} \rangle /
\langle \lambda_{1, \, \nu = 0} \rangle \,$. Our data are compared with
the theoretical prediction based on a combination of chiral RMT in 
the $\epsilon$-regime and
analytical formulae for $\Sigma (m)$ from Refs.\ \cite{HHI,Smilga}.}
%\end{center}
\label{EVratvsm}
\vspace*{-5mm}
\end{center}
\end{figure}

\section{Conclusions}

The overlap hypercube fermion has some computational overhead
compared to the standard overlap fermion, but a number of benefits:
better locality, approximate rotation symmetry, improved scaling
and the applicability of a simplified HMC force. 
The restriction to low polynomials
is particularly favourable for the numerical stability.

In our application to the 2-flavour Schwinger model 
on a $16 \times 16$ lattice at $\beta =5$ 
we obtain useful acceptance rates and reliable reversibility.
We cumulated statistics at masses $m=0.01,$ $0.03,$ $0.06,$
$0.09,$ $0.12,$ $0.18$ and $0.24 \, $
in the sectors of topological charge $| \nu | = 0$ and 1 .
We revealed a new type of microscopic Dirac spectrum,
which is not explored analytically. Nevertheless, 
by combining RMT formulae for the spectrum with analytical expressions
for $\Sigma$, we obtained a prediction for the mass dependence of the
ratio $\langle \lambda_{1, \, | \nu | = 1} \rangle /
\langle \lambda_{1, \, \nu = 0} \rangle \, $, which matches our numerical 
data at $m \leq 0.12$ impressively well.

\appendix

\vspace{-2mm}

\section{Testing the dynamical overlap-HF in QCD}

We also implemented the HMC algorithm for $D_{\rm ovHF}$ in QCD,
with the HF force which can be chirally corrected with Zolotarev
polynomials of any degree $p$. We display thermalisation histories
of the dynamical overlap-HF in QCD, on $8^{4}$ lattices with
polynomial degrees $p = 6$ and $8$. They are
applied to the HMC force in the spectral interval 
$[ b , \lambda_{\rm max}]$ with a lower bound $b=0.1$ or $0.05$
($\lambda_{\rm max}$ is the maximal Hermitian eigenvalue).
For the precision parameter in eq.\ (\ref{simple}) we chose
$\vep = 10^{-4} $, the trajectory length now amounts to
$ \tau = 1$, and the accept/reject step is kept on machine precision.

\begin{figure}
\begin{center}
\vspace*{-4mm}
\includegraphics[angle=270,width=.58\linewidth]{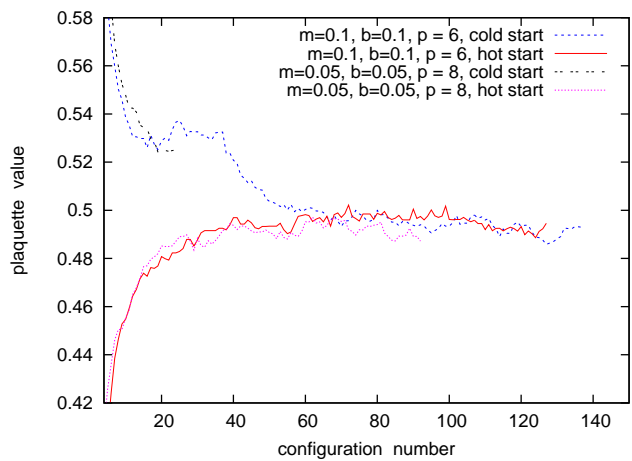}
\vspace*{-2mm}
\end{center}
\caption{A thermalisation history for the dynamical overlap-HF
in QCD at $\beta =5.6$ on a $8^{4}$ lattice. The algorithm is driven
by the full HF force --- corrected by a Zolotarev polynomial --- \
and the trajectory length is 1.}
\label{histo1}
\vspace*{-2mm}
\end{figure}

\begin{figure}
\begin{center}
\vspace*{-4mm}
\includegraphics[angle=270,width=.5\linewidth]{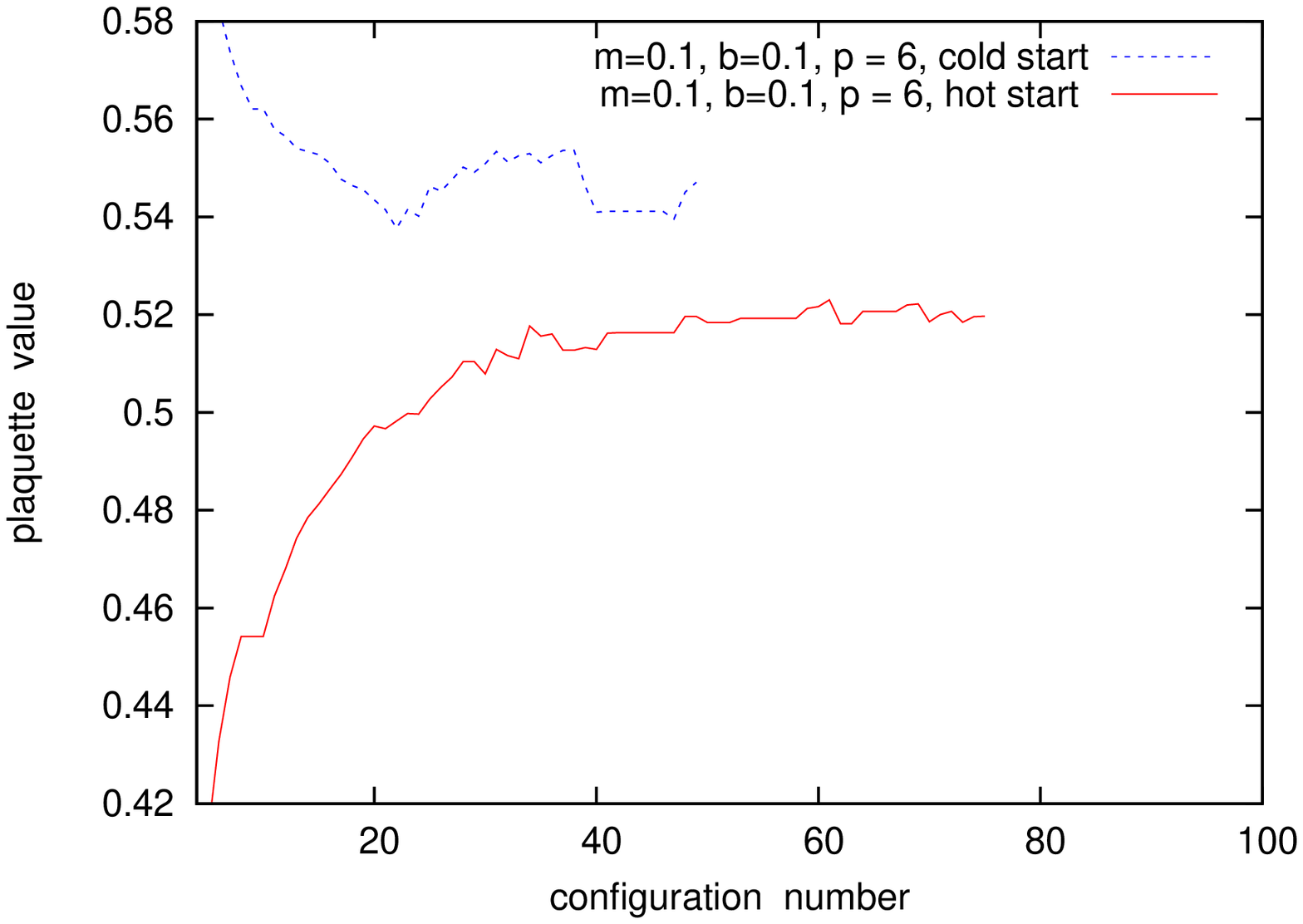}
\hspace*{-3mm}
\includegraphics[angle=270,width=.5\linewidth]{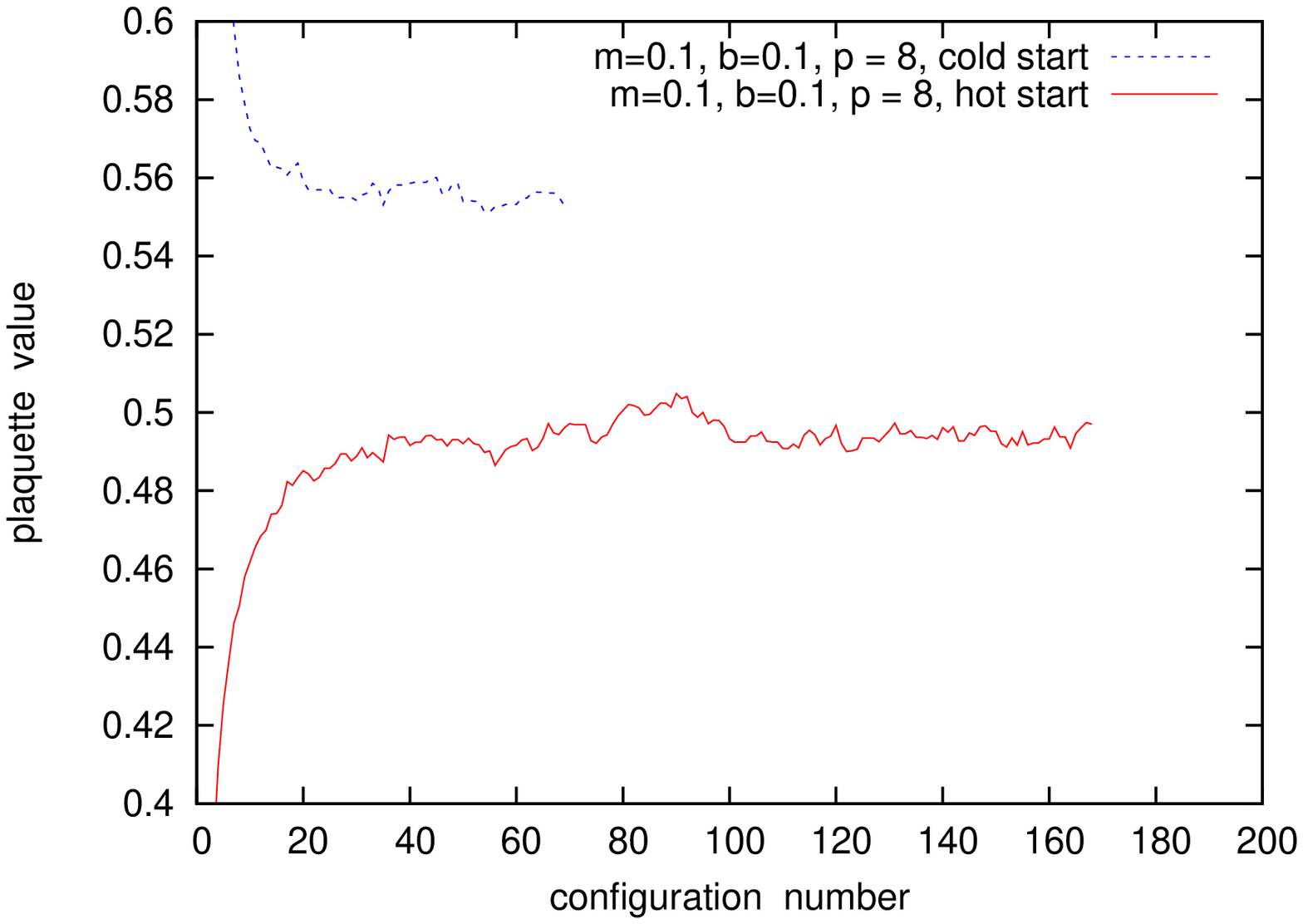}
\end{center}
\caption{Thermalisation history for the dynamical overlap-HF
in QCD on a $8^{4}$ lattice at $\beta =5.7$ (on the left) and
$\beta =5.8$ (on the right). In contrast to $\beta =5.6$
(Figure 5) the trajectories for cold and hot starts
level out on  different plateaux, which indicates a first
order phase transition.}
\label{histo23}
\end{figure}

At $\beta =5.6$ thermalisation sets in without problems (see
Figure \ref{histo1}),
whereas $\beta =5.7$ is plagued by a first order
phase transition, which is further pronounced
at $\beta =5.8$ (Figure \ref{histo23}). \\
%We plan applications in the $\epsilon$-regime
%of QCD, along the lines of Refs.\ \cite{epsQCD}. \\

\vspace*{-1mm}
\noindent
{\bf Acknowledgement :} We are indebted to Poul Damgaard,
Stephan D\"{u}rr, Martin Hasenbusch, Urs Heller, Jacques
Verbaarschot and Tilo Wettig for helpful discussions.
J.V.\ was supported by the ``Deutsche Forschungsgemeinschaft'' 
(DFG). The computations were performed on the %IBM 
p690 clusters of the ``Norddeutscher Verbund f\"ur Hoch- und 
H\"ochstleistungsrechnen'' (HLRN).

\end{document}